\documentclass[a4paper]{jpconf}
\usepackage{graphicx}
\usepackage{ulem} 	% unterstreichen
\usepackage{color}
\begin{document}
\title{Building blocks for future detectors: Silicon test masses and 1550\,nm laser light}

\author{R~Schnabel$^1$, M~Britzger$^1$, F~Br\"{u}ckner$^2$, O~Burmeister$^1$, K~Danzmann$^1$, J~D\"{u}ck$^1$, T~Eberle$^1$, D~Friedrich$^1$, H~L\"{u}ck$^1$, M~Mehmet$^1$, R~Nawrodt$^3$, S~Steinlechner$^1$, and B~Willke$^1$}

\address{$^1$ Albert-Einstein-Institut, Max-Planck-Institut f\"{u}r Gravitationsphysik, Institut f\"{u}r Gravitationsphysik der
Leibniz Universit\"{a}t Hannover, Callinstr. 38, 30167 Hannover,
Germany}

\address{$^2$ Institut f\"{u}r Angewandte Physik, Friedrich-Schiller-Universit\"{a}t Jena, Max-Wien-Platz 1, 07743
Jena, Germany}

\address{$^3$ Institute for Gravitational Research, University of Glasgow, G12 8QQ Glasgow, Scotland}

\ead{roman.schnabel@aei.mpg.de}

\begin{abstract}
Current interferometric gravitational wave detectors use the combination of quasi-monochromatic, continuous-wave laser light at 1064\,nm and fused silica test masses at room temperature. Detectors of the third generation, such as the Einstein-Telescope, will involve a considerable sensitivity increase. The combination of 1550\,nm laser radiation and crystalline silicon test masses at low temperatures might be important ingredients in order to achieve the sensitivity goal. Here we compare some properties of the fused silica and silicon test mass materials relevant for decreasing the thermal noise in future detectors as well as the recent technology achievements in the preparation of laser radiation at 1064\,nm and 1550\,nm relevant for decreasing the quantum noise. We conclude that silicon test masses and 1550\,nm laser light have the potential to form the future building blocks of gravitational wave detection. 
\end{abstract}

\section{Introduction}
The measuring principle of interferometric gravitational wave detectors is based upon laser beams that monitor changes in the distances between quasi-free falling test masses \cite{Sau1994,Row2000,Auf2005}. A high sensitivity to gravitational waves require a great number of cutting edge-technologies. The test masses need to be suspended as multiple pendulums in order to mechanically decouple them from the environment. They should have masses of several tens of kilogramms in order to reduce their susceptibility to radiation pressure forces and they should have a very high reflectivity for the laser wavelength used. The laser beams must show very low fluctuations in all their mode parameters like polarisation, spatial intensity distribution, frequency and power. In this contribution we focus on the thermal noise of the test masses, i.e. the thermally excited motions of the test mass surfaces with respect to the center of masses, and on the shot-noise, i.e. the photon counting noise of the photo-electric detection of the laser light. We are in particular interested in potential noise reductions that might be achievable if the current fused silica test mass material is replaced by crystalline silicon, as previously considered in Refs.~\cite{WDRS91,Rowan03,RHC05}, and the laser wavelength changed from 1064\,nm to 1550\,nm.

\section{Fused silica and crystalline silicon}
According to the equipartition theorem, thermal noise can be reduced by decreasing the temperature. Additionally, as decribed by the fluctuation-dissipation theorem \cite{Cal1951,Cal1952}, the thermal noise at Fourier frequencies far away from the system's mechanical resonances can be reduced by increasing the mechanical quality factors (Q-factors), if the mechanical dissipation associated with the sample under study is spatially homogeneous. The higher the Q-factor, the more thermal noise is transferred from the off-resonant spectrum into the resonances. The detection band of gravitational wave detectors is below the fundamental internal resonance frequency of test mass mirrors, which implies that a high Q-factor is very useful and therefore an essential requirement for detector test masses.
\begin{figure}[ht]
	\centering
		\includegraphics[width=11cm]{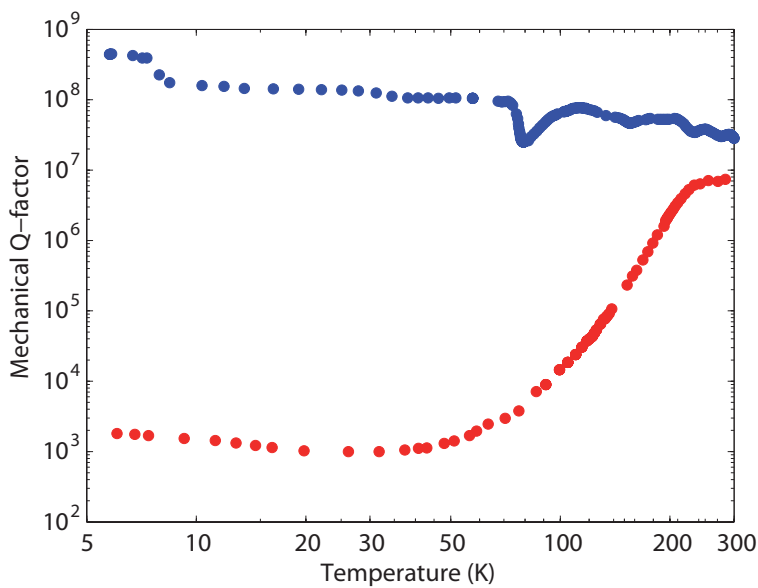}
		\caption{Mechanical quality factors of crystalline silicon (top, blue) and fused silica (bottom, red) versus temperature. The data were measured at the Friedrich-Schiller-Universit\"{a}t Jena on a 11.6\,kHz mode of a fused silica and on a 14.6\,kHz mode of a silicon(100) substrat. Both substrates were cylindrical, were 12\,mm thick and had a diameter of 3" \cite{SNHTNVTS09}.}
	\label{fig1:Q-factors}
\end{figure}

Current interferometric detectors are operated at room temperature with test masses made from fused silica. Fused silica offers high Q-factors of at least $2\times10^{8}$ at room temperature \cite{APDPS04}. However, future gravitational wave detectors require even higher Q-factors and, most likely, also an operation of the test masses at temperatures far below room temperature. Unfortunately, Q-factors of fused silica samples significantly decrease with decreasing temperatures \cite{MSi53}-\nocite{McS53,FVK54,ABo55,Tie1992,Vac2005}\cite{TABCDGVM07}. This effect is also clearly found in the measurement data of Fig.~\ref{fig1:Q-factors} (bottom trace, red dots) \cite{SNHTNVTS09,NZKSHHNTNVST08}. Crystalline silicon, on the other hand, generally shows high Q-factors that even increases with lower temperature (top trace, blue dots in Fig.~\ref{fig1:Q-factors}). A Q-factor of even more than 10$^9$ at cryogenic temperatures was observed in \cite{MLGHDG78}. The high Q-factors of silicon, which are also available at room temperature, resulted in proposals that this material might be a good test mass material for gravitational wave detectors as early as in 1991 \cite{WDRS91,Rowan03,RHC05}.\\
\begin{figure}[ht]
	\centering
		\includegraphics[width=10cm]{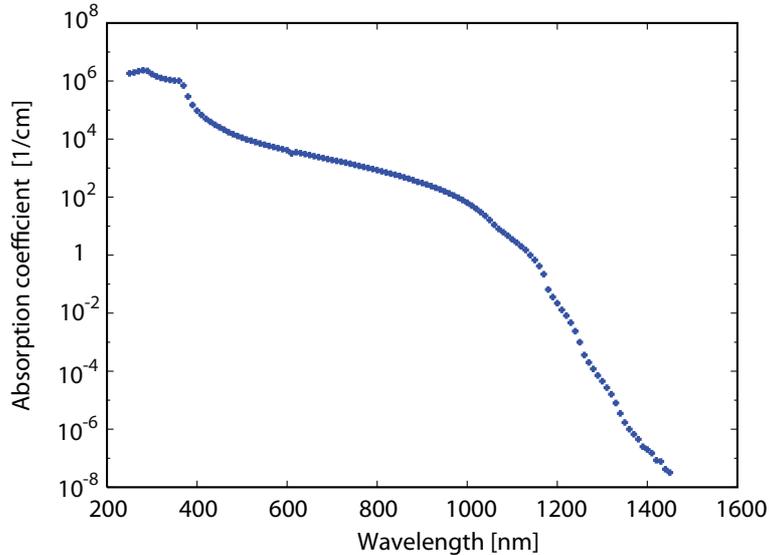}
		\caption{Measured optical absorption coefficients of silicon. The data, taken from Ref.~\cite{GKe95}, suggest an absorption of the order of 10$^{-8}$cm$^{-1}$ at wavelengths around 1550\,nm.}
	\label{fig2:Si-Absorption-Green}
\end{figure}

An obvious problem with silicon is its rather high optical absorption at the currently used laser wavelength of 1064\,nm. The absorption coefficient of silicon at 1064\,nm is about 10\,cm$^{-1}$ \cite{GKe95}, see Fig.~\ref{fig2:Si-Absorption-Green}. In contrast fused silica may show absorption at 1064\,nm as low as $2.5\times 10^{-7}$cm$^{-1}$, such as the Heraeus Suprasil\,3001. Test mass mirrors in gravitational wave detectors also serve as transmissive optical coupling mirrors to the Fabry-Perot arm resonators. Therefore, the test mass materials need to have extremely low absorption in order to minimize thermal lensing and photon absorption induced thermo-refractive noise \cite{BGV00}.
In order to be able to use non-transparent materials like silicon in combination with 1064\,nm radiation, all-reflective schemes have been proposed and demonstrated \cite{Dre95,SBy97,BBBDSCKT04,BBCKTDS06}. Such schemes are based on diffractive couplers and could, in principle, solve this problem. \\
However, a signficant and actually dominating source of thermal noise arises from the high-reflectivity dielectric multi-layer coating stacks on the test mass mirrors. Such coatings consist of a several micrometer thick stack of up to 40 alternating layers of low- and high-refractive index materials. Recent research has shown that in particular the typically used high-index material tantala (Ta$_2$O$_5$) is responsible for the dominant thermal noise contribution of the coating \cite{Penn03}. At a laser wavelength of 1064\,nm the coating thermal noise problem is very difficult to solve because of a lack of high index materials with high mechanical quality and low optical absorption at that wavelength. Doping of Ta$_2$O$_5$ with TiO$_2$ has also been investigated and a reduction of the mechanical loss by a factor of 1.5 was observed at room temperature \cite{harry2}. If a wavelength is used at which silicon is highly transparent, not only can the effects of thermal lensing and the photon absorption induced thermo-refractive noise be minimized, but also the coating thermal noise can be reduced. Since silicon has a rather high index of refraction of 3.48 it could serve as an ideal high-index material in a conventional dielectric multi-layer stack. A combination of silicon with silica would generate the same high reflectivity as a combination of tantala and silica but with less layers. One inch optics with such coating stacks are already successfully used in our experiments as steering mirrors. A precise measurement of the reflectivities achieved is in preparation. The data from Ref.~\cite{GKe95} as presented in Fig.~\ref{fig2:Si-Absorption-Green} shows that at wavelengths above 1064\,nm the absorption of silicon is drastically reduced. At the telecommunication wavelength of 1550\,nm the absorption might be below $10^{-8}$cm$^{-1}$. Since the measurement of such low absorptions is difficult reliable data in this wavelength region is still missing. \\
Recently, a coating-free, monolithic high-reflectivity surface of crystalline silicon was proposed \cite{BCBFKDTS08} and fabricated \cite{BFCBBDKTS09}. A reflectivity of 99.8\% was achieved for a silicon surface without adding any material by etching a specific structure into the crystalline silicon surface. In principle this method also works at other wavelengths and other materials. However, again the high refraction index of silicon is very important for designing an error-tolerant high-reflection surface and the combination of silicon and 1550\,nm seems to be an excellent choice.

Of course several parameters of a material are important to judge its suitability as a gravitational wave test mass mirror. This contribution does not aim for a comprehensive survey. Only some other promissing properties of silicon should be mentioned here. Silicon has a vanishing linear thermal expansion coefficient at temperatures of 125\,K and around 20\,K \cite{OTo84}. At these temperatures the thermo-elastic noise of silicon is therefore minimal.\\
Another positive feature of silicon is its rather high thermal conductivity of about 2\,W/(cm\,K) at room temperature. At 125\,K and at 20\,K the thermal conductivities are even higher, of about 10\,W/(cm\,K) \cite{GSl64}. These high values would help in cooling the test mass mirrors down to cryogenic temperatures and in avoiding the thermal lensing problem. The values for fused silica are considerably lower, 0.015\,W/(cm\,K) at room tempeature, less than 0.01\,W/(cm\,K) at 125\,K and 0.001\,W/(cm\,K) at 20\,K \cite{Dam73}.

\section{Laser wavelengths of 1064\,nm and 1550\,nm}
\begin{figure}[ht]
	\centering
		\includegraphics[width=11cm]{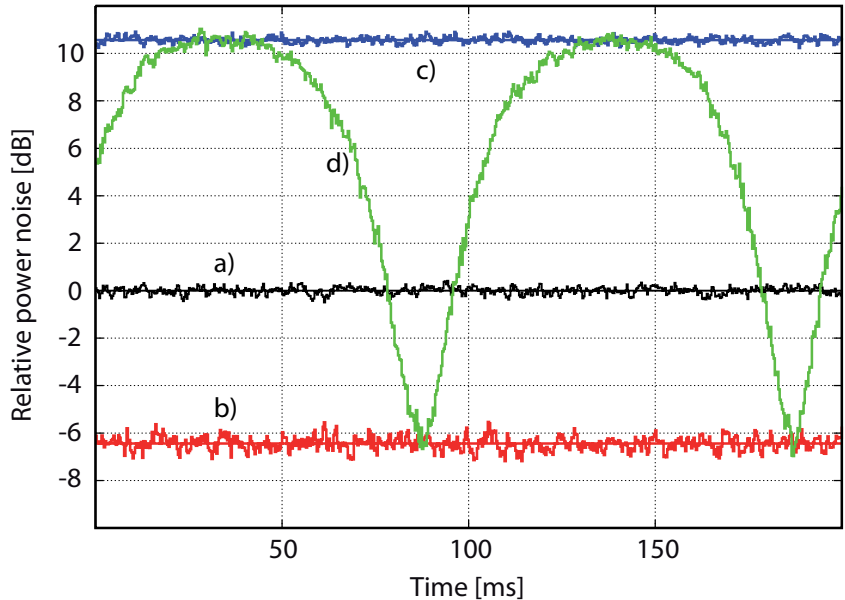}
		\caption{Noise powers at a Fourier frequency of 5\,MHz, normalized to the laser beam's shot noise as shown in trace (a). Trace (b) represents the observation of 6.4\,dB squeezed shot noise, and trace (c) the observed anti-squeezing. In both traces the phase angle of the squeezed light beam was controlled to be fixed. Trace (d) shows the noise power when the phase angle was scanned.}
	\label{fig3:SQZ1550}
\end{figure}

Interferometric gravitational wave detectors require ultra-stable high-power laser radiation in order to reach high measurement sensitivities. The higher the laser power, the lower is the photo-electric counting noise in relation to the measurement signal, i.e. the signal-normalized shot-noise. After many years of research and developement of solid-state laser-systems, now ultra-stable single-mode laser radiation of more than 200\,W at the wavelength of 1064\,nm is available \cite{FWKF05}. A laser-system of such power has been developed for the Advanced LIGO detector at the Laser Zentrum Hannover and the Albert-Einstein-Institut Hannover \cite{WDFKKKPSSSVWWW08}.  Systems with powers from 10\,W to 35\,W have been operated in current gravitational wave detectors and a lot of expertise at 1064\,nm is therefore available.\\
In view of future detector generations the laser wavelength of 1064\,nm also offers a rather mature technology for generating squeezed states of light. In addition to the high-power laser light input, future detector generations will, most likely, also have a (dim) squeezed light input into the (almost dark) signal out-put port of the interferometer. This additional light input reduces (`squeezes') the measurement's photon counting statistics. Most importantly, with squeezed light the shot-noise is reduced without a laser power increase, i.e. without a thermal load increase inside the interferometer. Since even the best mirror test mass materials have a residual absorption, this property is highly valuable for the operation at low temperatures. Suspended test masses are highly decoupled from the environment and heating due to light absorption is not compatible with a low temperature operation.\\ 
Squeezed light sources at 1064\,nm were tested in many table-top experiments and at the suspended 40m-prototype at the California Institute of Technology \cite{MSMBL02,VCHFDS05,Goda08NatPhys}. For the first time, squeezed shot-noise in the audio detection band of gravitational wave detectors was demonstrated \cite{MGBWGML04,VCDS07}, a detector compatible control scheme developed \cite{VCHFDS06} and the strongest squeezing effect ever observed \cite{Vahlb08,Mehmet11dB}.\\
At 1550\,nm, research and developement of laser light sources for gravitational wave detectors have just begun. Erbium-doped fibre lasers are commercially available from a number of companies and provide up to 2\,W of continuous-wave single-mode laser radiation around 1550\,nm. %Based on the expertise gained at 1064\,nm, single-mode output powers of the order of 100\,W at 1550\,nm should be achievable in the next years. 
Research and developement in the area of high-power fiber amplifiers is currently under way at the Laser Zentrum Hannover and the Albert-Einstein-Institut (AEI) supported by QUEST (Quantum Engineering and Space Time Research), the centre of excellence at the Leibniz Universit{\"a}t Hannover. We expect that single-mode output powers of the order of 100\,W should be achievable at 1550\,nm within the next years.\\
Strongly squeezed light at 1550\,nm was recently realized at the AEI Hannover \cite{MSEVTDS09}. An improved result, squeezing of 6.4\,dB below the shot-noise is presented in this contribution, see Fig.~\ref{fig3:SQZ1550}. Here, trace (a) represents the noise power (the variance) of the shot-noise of a laser beam with a certain power. Trace (b) shows the squeezed noise of a laser beam with the same power. Here the phase angle of the squeezed field quadrature is controlled, which is the setting one would use for a measurement sensitivity improvement. The set-up used to generate squeezed light at 1550\,nm is in direct analogy to the previously used set-ups at 1064\,nm. Starting from a few tenths of a Watt of laser radiation at the fundamental wavelength (here 1550\,nm) laser radiation at the second harmonic frequency is produced that subsequently pumps sub-threshold optical parametric oscillation in a second-order nonlinear crystal. The down-converted field, again at fundamental frequency, shows the squeezed photon statistics at sub-shot-noise levels. Clearly, the expertise gained at 1064\,nm can directly be transferred in order to optimize the squeezed light sources at 1550\,nm, eventually reaching the same performances already demonstrated at 1064\,nm.

\section{Conclusion}
The combination of silicon as the test mass material and 1550\,nm as the laser wavelength has a high potential to provide the building blocks of future generations of gravitational wave detectors like the European Einstein-Telescope \cite{ET}. Silicon has excellent well-known mechanical properties at low temperatures, in particular a high mechanical Q-factor, high thermal conductivity, and a vanishing thermal expansion coefficient at around 20\,K and 125\,K. Additionally, silicon most likely shows very low optical absorption at 1550\,nm of the order 10$^{-8}$cm$^{-1}$. The high index of refraction of silicon might be very useful to solve the coating noise problem, for wavelengths around 1550\,nm. A laser material for this wavelength region already exists, and the developement of ultra-stable high-power lasers and state of the art squeezed light sources at 1550\,nm should be possible. We also would like to mention here that the transition to a longer wavelength reduces Rayleigh scattering and relaxes tolerance requirements for the generation of high-reflectivity test mass surfaces. Last but not least, silicon industry as well as telecommunication industry already offer a great availability of ultra-pure silicon crystals and laser optical components at 1550\,nm, which certainly would ease a transition from fused silica and 1064\,nm to crystalline silicon and 1550\,nm  laser light.

\ack{This work was supported by the Centre for Quantum Engineering and Space-Time Research QUEST and the Deutsche Forschungsgemeinschaft (DFG) within the Sonderforschungsbereich TR7. The research leading to these results has also received funding from the European Community's Seventh Framework Programme (FP7/2007-2013) under grant agreement N. 211743 (Einstein Telescope Design Study).}


\section*{References}
\begin{thebibliography}{9}
% \bibitem{Sto2009} Stochino A {\it et al.} 2009 {\it Nucl Instrum Meth A} {\bf 598} 737

\bibitem{Sau1994} Saulson P R 1994 \textit{Fundamentals of Interferometric Gravitational Wave Detectors} (Singapore: World Scientific)

\bibitem{Row2000} Rowan S and Hough J 2000 %Gravitational wave detection by interferometry (ground and space) 
\textit{Living Rev. Relativity} \textbf{3} http://www.livingreviews.org/lrr-2000-3

\bibitem{Auf2005} Aufmuth P and Danzmann K 2005 %Gravitational wave detectors 
\textit{New J. Phys.} \textbf{7} 202 1%-15

\bibitem{Cal1951} Callen H B and Welton T A 1951 %Irreversibility and Generalized Noise 
\textit{Phys. Rev.} \textbf{83} 34%-40

\bibitem{Cal1952} Callen H B and Greene R F 1952 %On a Theorem of Irreversible Thermodynamics 
\textit{Phys. Rev.} \textbf{86} 702%-10

\bibitem{APDPS04} Ageev A, Palmer B C, De Felice A, Penn S D and Saulson P R 2004 % Very high quality factor measured in annealed fused silica
\textit{Class. Quantum Grav.} \textbf{21} 3887%–3892

\bibitem{Bra1999} Braginsky V B, Gorodetsky M L and Vyatchanin S P 1999 %Thermodynamical fluctuations and photo-thermal shot noise in gravitational wave antennae 
\textit{Phys. Lett. A} \textbf{264} 1%-10

\bibitem{MSi53}Marx J W and Sivertsen J.M. 1953 %Temperature dependence of the elastic moduli and internal friction of silica and glass
\textit{J. Appl. Phys.} \textbf{24} 81%-87.

\bibitem{McS53}McSkimin H J 1953 % Measurement of elastic constants at low temperatures by ultrasonic waves – data for silicon and germanium single crystals and for fused silica
\textit{J. Appl. Phys.} \textbf{24} 988%-997.

\bibitem{FVK54}Fine M E, Van Duyne H and Kenney N T 1954 % Low-temperature internal friction and elasticity effects in vitreous silica
\textit{J. Appl. Phys.} \textbf{25} 402%-405.

\bibitem{ABo55}Anderson O L and Bommel H E  1955 %Ultrasonic absorption in fused silica at low temperatures and high frequencies
\textit{J. Amer. Ceram. Soc.} \textbf{38} 125%-131.

\bibitem{Tie1992} Tielb\"{u}rger D, Merz R, Ehrenfels R and Hunklinger S 1992 %Thermally activated relaxation processes in vitreous silica: An investigation by Brillouin scattering at high pressures 
\textit{Phys. Rev. B} \textbf{45} 2750%-60

\bibitem{Vac2005} Vacher R, Courtens E and Foret M 2005 %Anharmonic versus relaxational sound damping in glasses. II. Vitreous silica 
\textit{Phys. Rev. B} \textbf{72} 214205 %1-11

%Eine weitere Referenz für niedrige Güten bei Fused Silica ist:
\bibitem{TABCDGVM07}Travasso F \textit{et al.} %, Amico P, Bosi L, Cottone F, Dari A, Gammaitoni L, Vocca H, Marchesoni F 
2007 \textit{Europhys. Lett.} \textbf{80} 50008

\bibitem{SNHTNVTS09}Schwarz C \textit{et al.} %, Nawrodt R, Heinert D, Th{\"u}rk M, Neubert R, Vodel W, T{\"u}nnermann A, Seidel P 
2009 % Cryogenic Setup for Q-factor measurements on bulk materials for future gravitational wave detectors, 
Proceedings of ICEC22-ICMC2008, edited by Chang H-M \textsl{et al.}, %The Korea Institute of Applied Superconductivity and Cryogenics
978-89-957138-2-2

\bibitem{NZKSHHNTNVST08}Nawrodt R \textit{et al.} %, A. Zimmer, T. Koettig, C. Schwarz, D. Heinert, M. Hudl, R. Neubert, M. Th{\"u}rk, S. Nietzsche, W. Vodel, P. Seidel, A. T{\"u}nnermann 
2008 \textit{Journal of Physics: Conference Series} \textbf{122}, 012008

\bibitem{MLGHDG78}McGuigan D F \textit{et al.} %, Lam C C, Gram R Q, Hoffman A W, Douglas D H and Gutche H W, 
1978 %Measurements of the mechanical Q of single-crystal silicon at low temperatures 
\textit{J. Low Temp. Phys.} \textbf{30} 621%-9

%Si als Detektormaterial wurde vorgeschlagen in (das sind die 3 Standard-Referenzen):
\bibitem{WDRS91}Winkler W, Danzmann K, R{\"u}diger A, Schilling R 1991, \textit{Phys. Rev. A} \textbf{44} 7022
%
\bibitem{Rowan03}Rowan S \textit{et al.} %Byer R L, Fejer M M, Route R, Cagnoli G, Crooks D R M, Hough J, Sneddon P H, Winkler W, 
2003 in: M. Cruise, P. Saulson (Eds.), \textit{Gravitational Wave Detection, Proceedings of SPIE}, SPIE, Wellingham, WA, p. 292

\bibitem{RHC05}Rowan S, Hough J, Crooks D 2005, \textit{Phys. Lett. A} \textbf{347} 25

\bibitem{GKe95}Green M and Keevers M 1995, \textit{Progress in Photovoltaic research \& Appl.} \textbf{3} 189

%\bibitem{Heraeus} http://optics.heraeus-quartzglas.com/de/productsapplications/productdetail 1938.aspx

%\bibitem{HLW06} Hild S, L\"{u}ck H, Winkler W, Strain K, Grote H, Smith J, Malec M, Hewitson M, Willke B, Hough J and Danzmann K 2006 %Measurement of a low-absorption sample of OH-reduced fused silica 
%\textit{Appl. Opt.} \textbf{45} 7269%-72

\bibitem{BGV00}Braginsky V B, Gorodetsky M L, and Vyatchanin S P 2000 %Thermo-refractive noise in gravitational wave antennae 
\textit{Phys. Lett. A} \textbf{271}, 303%–307 (2000)

\bibitem{Dre95}Drever R W P 1995, in \textit{Proceedings of the Seventh Marcel Grossman Meeting on General Relativity}, M. Keiser and
R. T. Jantzen, eds. (World Scientific)

\bibitem{SBy97}Sun K-X and Byer R L 1997 \textit{Opt. Lett.} \textbf{23}, 567

\bibitem{BBBDSCKT04} Bunkowski A, Burmeister O, Beyersdorf P, Danzmann K, Schnabel R, Clausnitzer T, Kley E B and
T{\"u}nnermann A 2004 \textit{Opt. Lett.} \textbf{29} 2342

\bibitem{BBCKTDS06} Bunkowski A \textit{et al.} %, Burmeister O, Clausnitzer T, Kley E-B, T{\"u}nnermann A, Danzmann K, and Schnabel R 
2006 %Optical characterization of ultrahigh diffraction efficiency gratings 
\textit{Appl. Opt.} \textbf{45} 5795

\bibitem{Penn03}Penn S D et al. 2003 %Mechanical loss in tantala/silica dielectric mirror coatings
\textit{Class. Quant. Grav.} \textbf{20} 2917%-2928.

%Schichtuntersuchungen bei  tiefen Temperaturen:
%\bibitem{MartinETAL08} Martin I W \textit{et al.} 2008 %Armandula H, Comtet C, Fejer M M, A. Gretarsson, G. Harry, J. Hough, J.-M. M. Mackowski, I. MacLaren, C. Michel, J.-L. Montorio, N. Morgado, R. Nawrodt, S. Penn, S. Reid, A. Remillieux, R. Route, S. Rowan, C. Schwarz, P. Seidel, W. Vodel, A. Zimmer, Measurements of a low-temperature mechanical dissipation peak in a single layer of Ta2O5 doped with TiO2,
%\textit{Class. Quantum Grav.} \textbf{25} 055005
%
%\bibitem{MartinETAL09} Martin I W \textit{et al.} 2009 %E. Chalkley, R. Nawrodt, H. Armandula, R. Bassiri, C. Comtet, M. M. Fejer, A. Gretarsson, G. Harry, D. Heinert, J. Hough, I. MacLaren, C. Michel, C. J.-L. Montorio, N. Morgado, S. Penn, S. Reid, R. Rout, S. Rowan, C. Schwarz, P. Seidel, W. Vodel, A. Woodcraft, The effect of TiO2 doping on dissipation in Ta2O5 coating layers for gravitational wave detectors, 
%\textit{Class. Quantum Grav.} \textbf{26} 155012

\bibitem{harry2} Harry G M \textit{et al.} 2006 %, H. Armandula, E. Black, D. R. M. Crooks, G. Cagnoli, J. Hough, P. Murray, S. Reid, S. Rowan, P. Sneddon, M. M. Fejer, R. Route, and S. D. Penn, ``Thermal noise from optical coatings in gravitational wave detectors  
\textit{Appl. Opt.} {\bf 45,} 1569%-1574

\bibitem{BCBFKDTS08}Br{\"u}ckner F \textit{et al.} %, Clausnitzer T, Burmeister O, Friedrich D, Kley E-B, Danzmann K, T{\"u}nnermann A, and Schnabel R 
2008 % Monolithic dielectric surfaces as new low-loss light-matter interfaces. 
\textit{Opt. Lett.} \textbf{33}, 264%-266

\bibitem{BFCBBDKTS09} Br{\"u}ckner F, Friedrich D, Clausnitzer T, Britzger M, Burmeister O, Danzmann K, Kley E-B, T{\"u}nnermann A and  Schnabel R 2009 %Realization of a monolithic high-reflectivity cavity mirror from a single silicon crystal. 
\textit{submitted} 

\bibitem{OTo84}Okada Y and Tokumaru Y J  1984 \textit{Appl. Phys.} \textbf{56} 2 314  

\bibitem{GSl64}Glassbrenner C J and Slack G A 1964 \textit{Phys. Rev.} \textbf{134} 4A A1058%-A1069. 

\bibitem{Dam73}Damon D H 1973 \textit{Phys. Rev. B} \textbf{8} 5860

\bibitem{WDFKKKPSSSVWWW08} Willke B \textit{et al.} 2008 %, K. Danzmann, Mike Frede, P. King, D. Kracht, P. Kwee, O. Puncken, J. R. L. Savage, B. Schulz, F. Seifert, C. Veltkamp, S. Wagner, P. Wessels, and L. Winkelmann, Stabilized lasers for advanced gravitational wave detectors, 
\textit{Class. Quantum Grav.} \textbf{25}, 114040

\bibitem{FWKF05} Frede M, Wilhelm R, Kracht D, and Fallnich C 2005 %Nd:YAG ring laser with 213 W linearly polarized fundamental mode output power, 
\textit{Opt. Express} \textbf{13} 7516

\bibitem{MSMBL02} McKenzie K, Shaddock D A, McClelland D E, Buchler B C, and Lam P K 2002 {\it Phys. Rev. Lett.} {\bf 88} 231102 

\bibitem{VCHFDS05} Vahlbruch H, Chelkowski S, Hage B, Franzen A, Danzmann K, and Schnabel R 2005 {\it Phys. Rev. Lett.} {\bf 95} 211102

\bibitem{Goda08NatPhys} Goda K \textit{et al.} %, O. Miyakawa, E. E. Mikhailov, S. Saraf, R. Adhikari, K. McKenzie, R. Ward, S. Vass, A. J. Weinstein, and N. Mavalvala, "A quantum-enhanced prototype gravitational-wave detector",
2008 {\it Nat. Phys.} {\bf 4} 472%-476

\bibitem{MGBWGML04} McKenzie K, Grosse N, Bowen W P, Whitcomb S E, Gray M B, McClelland D E, and Lam P K 2004 {\it Phys. Rev. Lett.} {\bf 93} 161105

\bibitem{VCDS07} Vahlbruch H, Chelkowski S, Danzmann K, and Schnabel R 2007 % Quantum engineering of squeezed states for quantum communication and metrology, 
\textit{New J. Phys.} \textbf{9} 371

\bibitem{VCHFDS06} Vahlbruch H \textit{et al.}  %, Chelkowski S, Hage B, Franzen A, Danzmann K, and Schnabel R 
2006 {\it Phys. Rev. Lett.} {\bf 97} 011101

\bibitem{Vahlb08}Vahlbruch H \textit{et al.}  % M. Mehmet, N. Lastzka, B. Hage, S. Chelkowski, A. Franzen, S. Gossler, K. Danzmann, and R. Schnabel Observation of Squeezed Light with 10 dB Quantum-Noise Reduction. 
2008 {\it Phys. Rev. Lett.} {\bf 100} 33602

\bibitem{Mehmet11dB} Mehmet M, Vahlbruch H, Lastzka N, Danzmann K, and Schnabel R. %Observation of squeezed states with strong photon number oscillations 
2009 \textit{submitted} arXiv:0909.5386

\bibitem{MSEVTDS09} Mehmet M \textit{et al.} %, Steinlechner S, Eberle S, Vahlbruch H, Th{\"u}ring A, Danzmann K, and Schnabel R 
2009 \textit{Opt. Lett.} \textbf{34} 7 % Observation of cw squeezed light at 1550 nm

\bibitem{ET} http://www.et-gw.eu/

%\bibitem{sapphire99} Uchiyama T, Tomaru T, Tobar M E, Tatsumi D, Miyoki S, Ohashi M, Kuroda K, Suzuki T, Sato N, Haruyama T, Yamamoto A, and Shintomi T 1999 Mechanical quality factor of a cryogenic sapphire test mass for gravitational wave detectors \textit{Phys. Lett. A} \textbf{261} 5-11
%\bibitem{Miy2006} Miyoki S \textit{et al} 2006 The CLIO project \textit{Class. Quantum Grav.} \textbf{23} S231-7

% \bibitem{Gos2009} Go{\ss}ler S {\it et al.} 2009 {\it submitted to Class Quantum Grav} 
% \bibitem{Sto2009} Stochino A {\it et al.} 2009 {\it Nucl Instrum Meth A} {\bf 598} 737


\end{thebibliography}
\end{document}